\documentclass[
    ,final            
  ]
  {aipproc}

\layoutstyle{8x11double}

\begin{document}

\title{Asymptotic silence in loop quantum cosmology}

\classification{98.80.Qc}
\keywords{asymptotic silence, loop quantum cosmology, BKL conjecture}

\author{Jakub Mielczarek}{
address={Institute of Physics, Jagiellonian University, Reymonta 4, 30-059 Cracow, Poland \\
Department of Fundamental Research, National Centre for Nuclear Research, 
Ho{\.z}a 69, 00-681 Warsaw, Poland}}

\begin{abstract}
The state of asymptotic silence, characterized by causal disconnection of the space points, 
emerges from various approaches aiming to describe gravitational phenomena in the limit of 
large curvatures. In particular, such behavior was anticipated by Belinsky, Khalatnikov and 
Lifshitz (BKL) in their famous conjecture put forward in the early seventies of the last century. 
While the BKL conjecture is based on purely classical considerations, one can expect 
that asymptotic silence should have its quantum counterpart at the level of a more fundamental 
theory of quantum gravity, which is the relevant description of gravitational phenomena in the 
limit of large energy densities. Here, we summarize some recent results which give support 
to such a possibility.  More precisely, we discuss occurrence of the asymptotic silence due to 
polymerization of space at the Planck scale, in the framework of loop quantum cosmology. 
In the discussed model, the state of asymptotic silence is realized at the energy density 
$\rho = \rho_c/2$, where $\rho_c$ is the maximal allowed energy density, being of the order 
of the Planck energy density. At energy densities $\rho > \rho_c/2$, the universe becomes 
4D Euclidean space without causal structure. Therefore, the asymptotic silence appears 
to be an intermediate state of space between the Lorentzian and Euclidean phases. 
\end{abstract}

\maketitle

\section{Introduction}

Asymptotic silence is a hypothetical state of space in which information between 
any two space-points cannot be exchanged. The concept of asymptotic silence is 
not a new one, and has appeared in literature under different names such as: Carrollian 
limit \cite{Levy}, anti-Newtonian limit \cite{Carlip:2012md} or ultralocality \cite{Isham:1975ur}.  

The state of asymptotic silence appears also in various contexts. Perhaps the best 
known is the so-called  Belinsky-Khalatnikov-Lifshitz (BKL) conjecture \cite{BKL}, 
which, in particular, states that close to a cosmological singularity, spatially separated points 
decouple. Such expectation was supported by analytical and numerical investigations. 
The predicted behavior is of purely classical origin and is a result of the nonlinear 
dynamics of GR.

\begin{figure}
\includegraphics[height=.15\textheight]{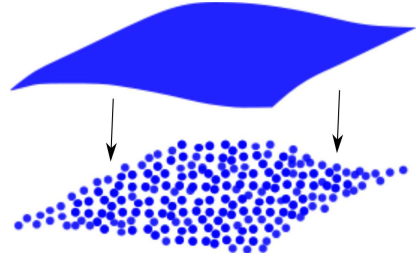}
\caption{Asymptotic silence can be seen as a process in which continuous space (top) 
decouple for a set of not interacting points (bottom). Information cannot propagate from 
one space point to another.}
\label{Silence}
\end{figure}

The pictorial representation of the asymptotic silence is shown in Fig. \ref{Silence}. 
In case of the asymptotic silence resulting from BKL scenario, each of the "points"
is described by the anisotropic cosmological solution.  

Besides the BKL scenario, the state of asymptotic silence can be recovered in
certain limits of physical constants. In particular, it is intuitively clear that silence 
is obtained while taking the speed of light $c\rightarrow 0$, which is known as the 
Carrollian limit. The less intuitive case is the strong coupling limit of the gravitational 
interactions, when $G \rightarrow \infty$. Relation between this limit and the asymptotic
silence can be understood by analyzing the Hamiltonian formalism of general relativity.
Namely, the scalar constraint can be schematically written as $S= G  \cdot \  kinetic + 
\frac{1}{G}\cdot \ potential$. Here, only the potential term contains spatial derivatives, 
which relate the neighboring points. Therefore, while talking $G \rightarrow \infty$ only 
the kinetic term survives, and the theory becomes ultralocal. 

An important property of the ultralocal limit is that the resulting algebra of constraints simplifies 
with respect to the classical one. Namely, the classical algebra of scalar constraint $S$ and 
spatial diffeomorphsims constraint $D$ is the following: 
\begin{eqnarray}
\left\{ D[N^a_1],D[N^a_2] \right\} &=& D[N_1^b\partial_bN^a_2-N_2^b\partial_bN^a_1], \nonumber \\
\left\{ S[N],D[N^a] \right\} &=& - S[ N^a \partial_a N],  \label{SSclass}  \\
\left\{ S[N_1],S[N_2] \right\} &=& s D \left[g^{ab}(N_1 \partial_b N_2 -N_2 \partial_b N_1)  \right], 
\nonumber
\end{eqnarray}
where $N$ is a lapse function, $N^a$ is a shift vector and $g^{ab}$ is the spatial metric.  Morever,  
$s=1$ corresponds to the Lorentzian signature and $s=-1$ to the Euclidean one. Due to the factor 
$g^{ab}$, the algebra of constraints is not a Lie algebra. 

In the ultralocal limit ($G\rightarrow \infty$) the algebra of constraints simplifies to the Lie algebra  
\begin{eqnarray}
\left\{ D[N^a_1],D[N^a_2] \right\} &=& D[N_1^b\partial_bN^a_2-N_2^b\partial_bN^a_1], \nonumber \\
\left\{ S[N],D[N^a] \right\} &=& - S[N^a \partial_a N ], \label{SSultra}  \\
\left\{ S[N_1],S[N_2] \right\} &=& 0.   
\nonumber
\end{eqnarray}
Surprisingly, the number of the local symmetry generators is the same as in GR. The same holds for 
Ho\v{r}ava-Lifshitz gravity in the $z \rightarrow0$ limit of the dynamical exponent \cite{Horava:2009uw}. 

\begin{figure}
\includegraphics[height=.17\textheight]{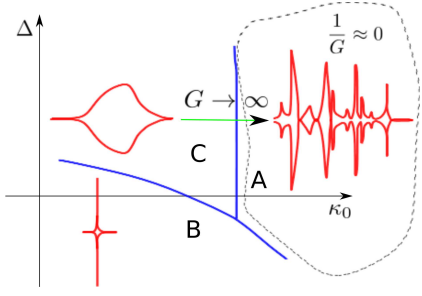}
\caption{Schematic illustration of the phase diagram of Causal Dynamical Triangulation.
Pase $A$, which corresponds  to the $G\rightarrow \infty$ limit, exhibits  some 
features of the asymptotic silence. For details see \emph{e.g.} Ref. \cite{Ambjorn:2012ij}.}
\label{CDT}
\end{figure}

Finally, recent results coming from the Causal Dynamical Triangulation (CDT) approach to 
quantum gravity suggest that a phase having properties of the asymptotic silence, 
appears just in the strong coupling limit. This regime corresponds to the phase $A$ on 
the phase diagram of CDT shown in Fig.  \ref{CDT}.
As observed from the numerical computations, Universe breaks up into several independent 
components, when the effective gravitational coupling constant G increases \cite{Ambjorn:2012ij}. 
This may suggest that in this phase the BKL scenario is realized. 

\section{Loop quantum cosmology}

How does the phase of asymptotic silence appear within the loop quantum cosmology?

At the effective level, the results of quantum polymerization of space, underlying 
loop quantum cosmology (LQC) \cite{Ashtekar:2011ni}, can be introduced by certain 
modifications of the classical constraints  $\mathcal{C}_{tot} \rightarrow \mathcal{C}^Q_{tot}$,
where $\mathcal{C}_{tot} = \mathcal{C}_{G}+\mathcal{C}_{M}$ is the total constraint
containing both, the gravity and matter part. In LQC, one usually considers two kinds of 
such quantum corrections: inverse volume corrections and holonomy corrections. 

It turns out, that not only the constraints, but also the algebra of constraints itself, is subject 
to modifications. One could thus say, that the particular form of the constraints 
enforces a deformation of the algebra of constraints.  Such deformations may have 
an anomalous form leading, due to lack of the closure of the algebra, to inconsistency of 
the formulation:  
\begin{equation}
\{ \mathcal{C}^Q_I, \mathcal{C}^Q_J \} = {g^K}_{IJ}(A^j_b,E^a_i) \mathcal{C}^Q_K+
\mathcal{A}_{IJ}.  
\end{equation}
The dynamics of the system preserves the surface of constraints only if the anomalous terms 
$\mathcal{A}_{IJ}$ are vanishing. Therefore, the $\mathcal{A}_{IJ} = 0$  restriction should be 
imposed. This is albeit very beneficial because it allows to remove ambiguities associated 
with introduction of the quantum corrections. Such procedure was successfully applied in case 
of the perturbative inhomogeneities on the FRW cosmological background.  

In case of scalar perturbations, the anomaly free formulation with inverse volume corrections, 
was found in Ref. \cite{Bojowald:2008gz}, whereas containing holonomy corrections, in Refs. 
\cite{Cailleteau:2011kr,WilsonEwing:2011es}. In what follows, we will focus on the results 
concerning the later. 

The results of \cite{Cailleteau:2011kr} indicate that there is a unique way of formulating the theory 
of scalar perturbations in the anomaly-free manner. Furthermore, anomaly freedom is fulfilled only 
if the physical area of the loop $\Delta$ is constant in time, which corresponds to the so-called \emph{new 
quantization scheme}. Up to the second order in the perturbative development, the resulting algebra 
of total constraints (which takes into account the scalar matter) is \cite{Cailleteau:2011kr} 
\begin{eqnarray}
\left\{ D_{tot}[N^a_1],D_{tot} [N^a_2] \right\} &=& 0, \nonumber \\
\left\{ S^Q_{tot}[N],D_{tot}[N^a] \right\} &=& - S^Q_{tot}[\delta N^a \partial_a \delta N],  \label{SSquant}  \\
\left\{ S^Q_{tot}[N_1],S^Q_{tot}[N_2] \right\} &=& \Omega D_{tot} \left[ \frac{\bar{N}}{\bar{p}} \partial^a(\delta N_2 - \delta N_1)\right].  
\label{HtotHtot} \nonumber
\end{eqnarray}
The first two terms of (\ref{SSquant}) agree with the classical expressions (\ref{SSclass}). The Poisson bracket 
of the two scalar constraints is however deformed with respect to the classical case, due to the presence of the 
factor 
\begin{equation}
\Omega=\cos(2\bar{\mu} \gamma\bar{k}) = 1 - 2\frac{\rho}{\rho_c} \in [-1,1], 
\end{equation}
where $\rho$ is the energy density of the matter field and the critical energy density
$\rho_c  := \frac{3}{8\pi G \Delta} \sim \rho_{Pl}$. Here, we assumed that the area of the 
loop $\Delta \sim l_{Pl}^2$, as expected from analysis of the area operator in loop quantum gravity. 

It is worth stressing that similar deformation of the algebra of constraints has been observed
also in different models. In particular, the same cosine type deformation 
was found for the spherically symmetrical models with holonomy corrections in LQC \cite{Reyes}. 
Deformation of the algebra of constraints has also similar form for the case of inverse volume 
corrections \cite{Bojowald:2008gz}. In that case however, the negative values of the deformation 
factor $\Omega$ are not allowed \cite{Bojowald:2011aa}.  Also, in case of the 2+1 loop quantum 
gravity, where the \emph{off-shell} calculations can be performed, similar deformation of the 
quantum algebra was found \cite{Perez:2010pm}. Therefore, there is accumulating evidence, 
that the obtained deformation of the algebra may be present also at the level of the full 3+1 
loop quantum gravity.

Let us focus now of the physical meaning of the obtained deformation of the algebra of constraints.  
In the limit of low energy densities $\rho/\rho_c \rightarrow 0$, the $\Omega \rightarrow 1$ recovering 
correctly the classical limit with signature $s=1$. However, while the energy density $\rho \rightarrow \rho_c/2$ 
the  $\Omega$ factor vanishes and the algebra of constraints (\ref{SSquant}) takes the ultralocal form (\ref{SSultra}). 
This  indicates that the state of asymptotic silence is realized at the energy densities $\rho = \rho_c/2$,  
which can be also interpreted as realization of the BKL conjecture \cite{Bojowald:2012qq}. Furthermore, 
the $\Omega$ factor becomes negative in the energy density range $(\rho_c/2, \rho_c]$. This region can 
be interpreted as the Euclidean phase, in contrast to the Lorentzian phase for $\rho < \rho_c/2$. For the 
maximal allowed energy density $\rho= \rho_c$, the algebra (\ref{SSquant}) simplifies to the classical one
(\ref{SSclass}), with the Euclidean signature $s=-1$. In order to clarify this issue, let us consider the 
resulting equation of motion for the gauge invariant Mukhanov variable $v$ \cite{Cailleteau:2011kr,Cailleteau:2012fy}:
\begin{eqnarray}
\frac{d^2}{d\eta^2}{v}-c^2_{eff} \nabla^2 v - \frac{z^{''}}{z} v = 0, 
\label{perteq}
\end{eqnarray}
where the effective speed of propagation: 
\begin{equation}
c_{eff} = \sqrt{\Omega} = \sqrt{1-2\frac{\rho}{\rho_c}}. 
\end{equation}

\begin{figure}
\includegraphics[height=.17\textheight]{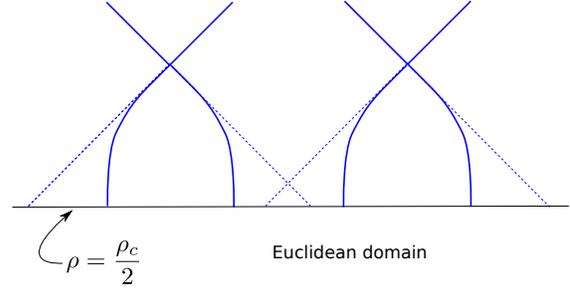}
\caption{Light cones for the model with deformed algebra of constraints. The asymptotic silence ($c_{eff} \rightarrow 0$) 
is realized while $\rho \rightarrow \frac{\rho_c}{2}$. For $\rho > \frac{\rho_c}{2}$ the space-time becomes Euclidean.}
\label{Cones}
\end{figure}

For the energy densities  $\rho < \rho_c/2$, the $c^2_{eff}>0$ and the equation is of the hyperbolic form.
However, for  the energy densities  $\rho > \rho_c/2$, the $c^2_{eff}<0$ and  equation becomes of the 
elliptic type. The same effect can be obtained by performing the Wick rotation, which changes the 
metric signature from Lorentzian to Euclidean. Based on this, we infer that the algebra of constraints
is deformed in such a way that, in the regime of high energies, the space-time becomes Euclidean. This 
process is visualized in Fig. \ref{Cones}, where the light cones for the discussed model are shown.  
It is worth stressing that the ultralocal limit of the algebra of constraints $\Omega \rightarrow 0$ translates to the 
Carrollian limit of the effective speed of light $c_{eff} \rightarrow 0$, in agreement with our expectations 
concerning the state of asymptotic silence.  
  
The state of the asymptotic silence, for which  $\rho = \frac{\rho_c}{2}$,  seems to be the right place to impose the 
initial conditions for the equations of motion. This is because it separates two regimes where equations of motion 
are of the different types.  One could expect that in the Euclidean regime the Dirichlet  or von Neumann boundary
conditions are imposed, while in the Lorentzian regime the Cauchy initial conditions are set at  $\rho = \frac{\rho_c}{2}$. 
Interestingly, in the vicinity of this point ($c^2_{eff} \approx 0$), the solution to equation 
(\ref{perteq}) has the following form 
\begin{equation}
v = z\left(A+B\int^{\eta} \frac{d\eta'}{z^2(\eta')}\right).
\end{equation}
The constants of integration $A$ and $B$ may be fixed \emph{e.g.} by imposing the \emph{silent initial conditions}
characterized by the lack of correlations between the neighboring space points, as expected for the
state of asymptotic silence. This would result in the white noise spectrum of initial perturbations.   

\section{Nature of the Euclidean phase}

Based one above results one can speculate that the spacetime is fundamentally Euclidean, 
and the Lorentzian structure emerges only in the limit of the low energy densities. In the high 
energy state ($\rho>\rho_c/2$), spacetime has no causal structure and the spatial and time 
directions are indistinguishable. Such a picture was, in fact, anticipated already in the early 
eighties of the last century by Hartle and Hawking in their famous no-boundary proposal 
\cite{Hartle:1983ai}. The no-boundary proposal is based on the assumption that the Wick 
rotation gains the physical meaning at the Planck epoch. Our model replaces the sudden 
Wick rotation by the smooth change of the signature, with the intermediate state of the 
asymptotic silence. Such transition reminds phase transitions observed in solid states, 
and it is tempting to investigate such analogy in more details. 

Let us focus on the model with the metamaterials composed of magnetized nanowires, 
which behave as spins \cite{Mielczarek:2012pf}. Such nanowires are randomly 
orientated at high temperatures. While passing to the low temperatures, however, the 
wires align in a certain direction, spontaneously breaking the $SO(3)$ symmetry of the system.
The order parameter for the symmetry broken $SO(2)$ phase is the local magnetization 
and the corresponding Goldstone bosons manifest as spin waves. Interestingly, in the direction 
of magnetization, the dielectric permittivity becomes negative leading to the emergence of 
the new effective time variable \cite{Smolyaninov:2010wi}. So, at the level of the equations 
of motion for the electric field propagating in the considered material, the original $SO(3)$ symmetry
of Laplace operator is replaced by $SO(1,2)$. Furthermore, it is worth stressing that below the 
cut-off scale, propagation of waves is not affected by the structure of the medium.   

In case of gravity, we observe that the symmetry of equations for the fields change from $SO(4)$ 
in the Euclidean region to $SO(1,3)$ in the Lorentzian regime. Taking the above solid state
model into account, one could speculate that this transition is a result of the symmetry breaking at 
the level of the fundamental structure of spacetime. In particular, one can suppose that the original 
$SO(4)$ spacetime symmetry is broken into $SO(3)$, where the residue $SO(3)$ is the rotational 
symmetry of triads. The time direction can be therefore seen as the order parameter of the 
symmetry broken phase. Interestingly, in such a picture, Goldstone bosons associated with the 
broken symmetry must appear. Such particles could naturally serve as inflatons, which are required 
to explain the inflationary stage after the Planck epoch. In our case, the Goldstone bosons 
are produced just at the end of the Planck epoch ($\rho=\rho_c/2$), triggering the subsequent 
inflationary stage. This makes the above hypothesis very promising and awaiting for a concrete
mathematical implementation. 

\section{Conclusions}

Asymptotic silence is an intriguing theoretical concept which may describe the state of spacetime 
in the limit of strong curvatures. It appears in various approaches to quantum gravity and therefore, it 
seems to be the ideal place to study relations between different formulations, such as loop quantum gravity, 
causal dynamical triangulation and Ho\v{r}ava-Lifshitz gravity. 

It was shown that, in loop quantum cosmology, the asymptotic silence appears in result of 
deformation of the algebra of constraints. The phase of asymptotic silence separates the Lorentzian
low curvature regime ($\rho<\rho_c/2$) from the high curvature Euclidean domain ($\rho>\rho_c/2$).
At $\rho=\rho_c/2$, the  effective speed of light falls to zero, which gives possible relation with theories 
of varying physical constants.  

The presented results contribute to the growing evidence, showing that the asymptotic silence
may indeed have something to do with the state of spacetime under very extreme conditions.
However, the final vote belongs to experiment. Assuming the silent initial conditions for cosmological 
perturbations at $\rho=\rho_c/2$, predictions regarding the anisotropies and polarization of the CMB 
can be performed. However, while such analysis is relatively straightforward to do, the potentially 
measurable effects typically occur at the length scales much greater than the present size of 
the cosmological horizon, and therefore are unaccessible observationally. The possible modifications 
of the photon dispersion relation would be perhaps easier to probe experimentally. Namely, since the
algebra of constraints is deformed, the corresponding Poincar\'{e} algebra should be deformed as well. 
In particular, we can expect that the Poincar\'{e} group $ISO(1,3)$ reduces to the Euclidean $E(4)$ 
group in the high energy limit, with  the intermediate stage described by the Carrollian group. Such 
deformation is conceptually similar with doubly special relativity and the $\kappa$-Poincar\'{e} 
deformation \cite{Lukierski:1991pn}. The dispersion relation of photons (as well as other particles) 
should be therefore subject to modifications, leading to time lags of the gamma ray bursts.

We hope that the presented results and ideas will stimulate further research on the issue of the 
asymptotic silence in quantum gravity.   
  
\bibliographystyle{aipproc}

\end{document}